\documentclass[conference]{IEEEtran}
\usepackage{naesseth}
\usepackage{abbreviations}

\ifCLASSINFOpdf
\else
\fi

\usepackage{algorithm}
\usepackage{algorithmic}

\usepackage[caption=false,font=footnotesize]{subfig}
\usepackage{tikz}
\usetikzlibrary{matrix,chains,scopes,positioning,arrows,fit}
\usepackage{color}
\usepackage[textwidth=2cm, textsize=scriptsize]{todonotes}

\begin{document}
%\title{Capacity estimation of two-dimensional models using Sequential Monte Carlo}
\title{Capacity estimation of two-dimensional channels using Sequential Monte Carlo}

\author{\IEEEauthorblockN{Christian A. Naesseth}
\IEEEauthorblockA{Division of Automatic Control\\
Link\"oping University\\
Link\"oping, Sweden\\
Email: christian.a.naesseth@liu.se}
\and
\IEEEauthorblockN{Fredrik Lindsten}
\IEEEauthorblockA{Department of Engineering\\
University of Cambridge\\
Cambridge, United Kingdom\\
Email: fredrik.lindsten@eng.cam.ac.uk}
\and
\IEEEauthorblockN{Thomas B. Sch\"on}
\IEEEauthorblockA{Department of Information Technology \\
Uppsala University\\
Uppsala, Sweden\\
Email: thomas.schon@it.uu.se}}


%Telephone: (800) 555--1212, Fax: (888) 555--1212}


\maketitle

\begin{abstract}
%\boldmath
  We derive a new Sequential-Monte-Carlo-based algorithm to estimate the capacity of two-dimensional
  channel models. The focus is on computing the noiseless capacity of the $2$-D $(1, \infty)$
  run-length limited constrained channel, but the underlying idea is generally applicable. The
  proposed algorithm is profiled against a state-of-the-art method, yielding more than an order of magnitude improvement in
  estimation accuracy for a given computation time.
\end{abstract}


% \fi
\IEEEpeerreviewmaketitle

\section{Introduction}\label{sec:intro}
With ever increasing demands on storage system capacity and reliability there has been increasing interest in page-oriented storage solutions. For these types of systems variations of two-dimensional constraints can be imposed to help with, amongst other things, timing control and reduced intersymbol interference \cite{immink2004codes}. This has sparked an interest in analyzing information theoretic properties of two-dimensional channel models
%for use in \eg holographic data storage solutions \cite{siegel2006information}.
for use in \eg holographic data storage \cite{siegel2006information}.

Our main contribution is a new algorithm, based on sequential Monte Carlo (\smc) methods, for numerically estimating the capacity of two-dimensional channels. We show how we can utilize structure in the model to sample the auxiliary target distributions in the \smc algorithm exactly. The focus in this paper is on computing the noiseless capacity of constrained finite-size two-dimensional models. However, the proposed algorithm works also for various generalizations and noisy channel models.

Recently, several approaches have been proposed to solve the capacity estimation problem in two-dimensional constrained channels. These methods rely either on variational approximations \cite{sabato2012generalized} or on Markov chain Monte Carlo \cite{loeliger2009estimating, molkaraie2013information}. Compared to these methods our algorithm is fundamentally different; samples are drawn sequentially from a sequence of probability distributions of increasing dimensions using \smc coupled with a finite state-space forward-backward procedure. We compare our proposed algorithm to a state-of-the-art Monte Carlo estimation algorithm proposed in \cite{loeliger2009estimating, molkaraie2013information}. Using \smc algorithms has earlier been proposed to compute the information rate of one-dimensional continuous channel models with memory \cite{dauwels2008computation}. Although both approaches are based on \smc, the methods, implementation and goals are very different.

\section{Two-dimensional channel models}\label{sec:background}
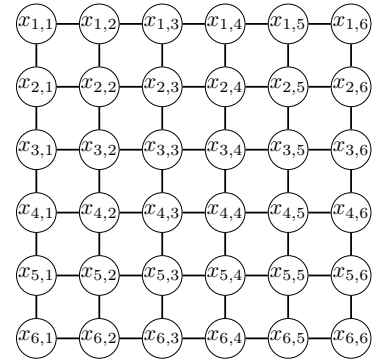
\begin{figure}[h!]
  \centering
  \resizebox{0.29\textwidth}{!}{%
    \tikzstyle{edge} = [-,thick]
\tikzstyle{arrw} = [very thick,shorten <=2pt,shorten >=2pt]
\tikzstyle{var} = [draw,circle,inner sep=0,minimum width=0.5cm]
\tikzstyle{obs} = [draw,circle,inner sep=0,minimum width=0.5cm, fill=black!20]
  \begin{tikzpicture}[>=stealth,node distance=0.6cm]
    \begin{scope}
      % Draw x-nodes and observations
      \foreach \x in {0,1,2,3,4,5} {
        \foreach \y in {0,1,2,3,4,5} {
          \pgfmathtruncatemacro\xend{\x+1}
          \pgfmathtruncatemacro\yend{6-\y}
          \node at (\x,\y) (x\x\y) [var] {$x_{\yend,\xend}$};
        }
      }
      % Draw horizontal edges
      \foreach \x in {0,1,2,3,4} {
        \pgfmathtruncatemacro\xend{\x+1}
        \foreach \y in {0,1,2,3,4,5} {
          \draw[edge] (x\x\y) -- (x\xend\y);
        }
      }
      % Draw vertical edges
      \foreach \x in {0,1,2,3,4,5} {
        \foreach \y in {0,1,2,3,4} {
          \pgfmathtruncatemacro\yend{\y+1}
          \draw[edge] (x\x\y) -- (x\x\yend) {};
        }
      }
    \end{scope}
     %\draw [blue] (current bounding box.south west) rectangle (current bounding box.north east);
  \end{tikzpicture}
  }
  \caption{$M \times M$ square lattice graphical model with pair-wise interactions. The nodes
    correspond to random variables $x_{\ell,j}$ and the edges encodes the interactions $\psi
    (x_{\ell,j},x_{m,n} )$.}\label{fig:pgm}
\end{figure}

As in \cite{molkaraie2013information} we consider the $2$-D $(1,\infty)$ run-length limited
constrained channel. The $2$-D $(1,\infty)$ run-length limited constraint implies that no two
horizontally or vertically adjacent bits on a $2$-D lattice may be both be equal to $1$. 
An example is given below:
\begin{align}
  {\footnotesize
  \begin{array}{ccccc}
    \cdots & \cdots & \cdots & \cdots & \cdots \\
    \cdots & 0 & 1 & 0 & \cdots \\
    \cdots & 0 & 0 & 1 & \cdots \\
    \cdots & 0 & 1 & 0 & \cdots \\
    \cdots & \cdots & \cdots & \cdots & \cdots
  \end{array}\nonumber}
\end{align}
This channel can be modelled as a probabilistic graphical model (PGM). A PGM is a probabilistic
model which {\em factorizes} according to the structure of an underlying graph $\G = \{\V, \E\}$,
with vertex set $\V$ and edge set $\E$. In this article we will focus on square lattice graphical
models with pair-wise interactions, see Figure~\ref{fig:pgm}.
That means that the joint probability mass function (\pmf) of the set of random variables, $\xV
\eqdef \{x_{1,1},\ldots,x_{1,M},x_{2,M},\ldots,x_{M,M} \}$, can be represented as a product of
factors over the pairs of variables in the graph:
\begin{equation}
  p( \xV ) = \frac{1}{Z} \prod_{(\ell j,m n) \in \E} \psi (x_{\ell, j}, x_{m, n}).
\label{eq:pfactors}
\end{equation}
Here, $Z$---the partition function---is given by
\begin{equation}
  Z = \sum_{\xV} \prod_{(\ell j,m n) \in \E} \psi (x_{\ell, j}, x_{m, n}), 
  \label{eq:Z}
\end{equation}
and $\psi (x_{\ell, j}, x_{m, n})$ denotes the so-called potential function encoding the pairwise
interaction between $x_{\ell, j}$ and $x_{m, n}$. For a more in-depth exposition of graphical models
we refer the reader to \cite{koller2009probabilistic}.

\subsection{Constrained channels and \pgm}
The noiseless $2$-D $(1,\infty)$ run-length limited constrained channel can be described by a square
lattice graphical model as in Figure~\ref{fig:pgm}, with binary variables $x_{\ell,j} \in \{0,1\}$
and pair-wise interactions between adjacent variables. Defining the factors as
\begin{equation}
  \psi (x_{\ell,j}, x_{m,n}) =
    \begin{cases}
      0, & \text{if } x_{\ell,j} = x_{m,n} = 1, \\
      1, & \text{otherwise,}
    \end{cases}
    \label{eq:interaction}
\end{equation}
results in a joint \pmf given by
\begin{align}
  p( \xV ) = \frac{1}{Z} \prod_{(\ell j,m n) \in \E} \psi (x_{\ell, j}, x_{m, n}),
\end{align}
where the partition function $Z$ is the number of satisfying configurations or, equivalently, the cardinality of the support of $p(\xV)$. For a channel of dimension $M \times M$ we can write the finite-size noiseless capacity as
\begin{align}
  \label{eq:bkg:capacity}
  C_M = \frac{1}{M^2} \log_2{Z}.
\end{align}
Hence, to compute the capacity of the channel we need to compute the partition function
$Z$. Unfortunately, calculating $Z$ exactly is in general computationally intractable. 
This means that we need a way to approximate the partition function. Note
that for this particular model, known upper and lower bounds of the infinite-size noiseless capacity,
$M \to \infty$, agree on more than eight decimal digits \cite{kato1999capacity,
  nagy2000capacity}. However, our proposed method is applicable in the finite-size case, as well as to other models where no tight
bounds are known.

\subsection{High-dimensional undirected chains}\label{sec:hmm}
In the previous section we described how we can calculate the noiseless capacity for $2$-D channel models by casting the problem as a partition function estimation problem in the \pgm framework. In our running example the corresponding graph is the $M \times M$ square lattice \pgm
depicted in Figure~\ref{fig:pgm}. We now show how we can turn these models into high-dimensional undirected chains
by introducing a specific new set of variables. We will see that this idea, although simple, is a
key enabler of our proposed algorithm.

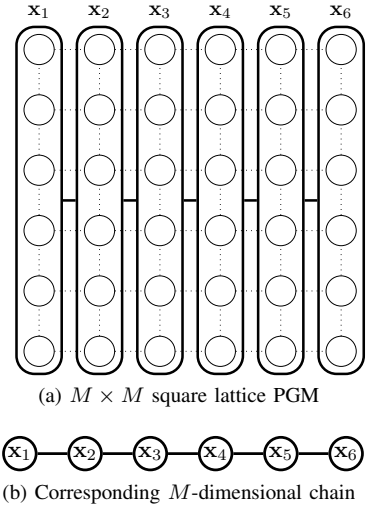
\begin{figure}[h!]
  \centering
  \subfloat[$M \times M$ square lattice \pgm]{
    \resizebox{0.29\textwidth}{!}{%
      \tikzstyle{edge} = [dotted]
\tikzstyle{edge2} = [-,very thick]
\tikzstyle{arrw} = [very thick,shorten <=2pt,shorten >=2pt]
\tikzstyle{var} = [draw,circle,inner sep=0,minimum width=0.5cm]
\tikzstyle{obs} = [draw,circle,inner sep=0,minimum width=0.5cm, fill=black!20]
  \begin{tikzpicture}[>=stealth,node distance=0.6cm]
    \begin{scope}
      % Draw x-nodes and observations
      \foreach \x in {0,1,2,3,4,5} {
        \foreach \y in {0,1,2,3,4,5} {
          \node at (\x,\y) (x\x\y) [var] {};
        }
      }
        
      \foreach \x in {0,1,2,3,4,5} {
        \pgfmathtruncatemacro\xend{\x+1}
        \node[draw,very thick,rectangle,rounded corners=3mm,minimum width=0.6cm,fit=(x\x0) (x\x5),label=above:{$\mathbf{x}_{\xend}$}] (x\x){};
      }
       Draw horizontal edges
      \foreach \x in {0,1,2,3,4} {
        \pgfmathtruncatemacro\xend{\x+1}
          \draw[edge2] (x\x) -- (x\xend);
      }
      % Draw horizontal edges
      \foreach \x in {0,1,2,3,4} {
        \pgfmathtruncatemacro\xend{\x+1}
        \foreach \y in {0,1,2,3,4,5} {
          \draw[edge] (x\x\y) -- (x\xend\y);
        }
      }
      % Draw vertical edges
      \foreach \x in {0,1,2,3,4,5} {
        \foreach \y in {0,1,2,3,4} {
          \pgfmathtruncatemacro\yend{\y+1}
          \draw[edge] (x\x\y) -- (x\x\yend) {};
        }
      }
    \end{scope}
     %\draw [blue] (current bounding box.south west) rectangle (current bounding box.north east);
  \end{tikzpicture}
      \label{fig:hmmfig_a}
    }
  }\\
  \subfloat[Corresponding $M$-dimensional chain]{
    \resizebox{0.29\textwidth}{!}{%
      \tikzstyle{edge} = [-,very thick]
\tikzstyle{arrw} = [very thick,shorten <=2pt,shorten >=2pt]
\tikzstyle{var} = [draw,circle,very thick,inner sep=0,minimum width=0.5cm]
  \begin{tikzpicture}[>=stealth,node distance=0.6cm]
    \begin{scope}
      % Draw x-nodes and observations
      \foreach \x in {0,1,2,3,4,5} {
        \foreach \y in {0} {
          \pgfmathtruncatemacro\xend{\x+1}
          \node at (\x,\y) (x\x\y) [var] {$\mathbf{x}_{\xend}$};
        }
      }
      % Draw horizontal edges
      \foreach \x in {0,1,2,3,4} {
        \pgfmathtruncatemacro\xend{\x+1}
        \foreach \y in {0} {
          \draw[edge] (x\x\y) -- (x\xend\y);
        }
      }

    \end{scope}
     %\draw [blue] (current bounding box.south west) rectangle (current bounding box.north east);
  \end{tikzpicture}
      \label{fig:hmmfig_b}
    }
  }
  \caption{$M \times M$ square lattice graphical model converted to an $M$-dimensional undirected chain model.}\label{fig:hmm}
\end{figure}

We define
 $\x_{k}$ to be the $M$-dimensional variable corresponding to all the original variables in column $k$, \ie 
\begin{align}
\x_{k} = \{ x_{1,k},\ldots,x_{M,k} \}, \qquad k=1,\ldots,M.
\end{align}
The resulting graphical model in the $\x_{k}$'s will be an undirected chain with joint \pmf given by
\begin{align}
p(\xV) = \frac{1}{Z} \prod_{k=1}^M \bphi (\x_{k}) \prod_{k=2}^M \bpsi( \x_{k}, \x_{k-1}),
\end{align}
where the partition function $Z$ is the same as for the original model and the $\bphi (\x_{k})$'s and $\bpsi( \x_{k}, \x_{k-1})$'s are the in-column and between-column interaction potentials, respectively. In terms of the original factors of the $2$-D $(1,\infty)$ run-length limited constrained channel model we get
\begin{subequations}
  \label{eq:psiDef}
  \begin{align}
    \bphi (\x_{k}) &= \prod_{j=1}^{M-1} \psi(x_{j+1,k}, x_{j,k}), \label{eq:inrow}\\
    \bm{\psi}( \x_{k}, \x_{k-1}) &= \prod_{j = 1}^M \psi (x_{j,k}, x_{j,k-1}). \label{eq:betweenrow}
  \end{align}
\end{subequations}
We illustrate this choice of auxiliary variables and the resulting undirected chain in Figure~\ref{fig:hmm}.
This transformation of the \pgm is a key enabler for the partition function estimation algorithm we propose in the subsequent section.

\section{Sequential Monte Carlo}\label{sec:smc}
Sequential Monte Carlo methods, also known as particle filters, are designed to sample sequentially from some sequence of target
distributions: $\bar{\gamma}_k(\x_{1:k})$, $k = 1,\,2\,\dots$. While SMC is most commonly used for inference on directed chains, in particular for state-space
models, these methods are in fact much more generally applicable. Specifically, as we shall see below,
SMC can be used to simulate from the joint \pmf specified by an undirected chain.
Consequently, by using the representation introduced in Section~\ref{sec:background}
it is possible to apply SMC to estimate the partition function of the 
$2$-D $(1,\infty)$ run-length limited constrained channel.
We start this section with a short introduction to \smc
samplers with some known theoretical results. These results are then used to compute an unbiased
estimate of the partition function. We leverage the undirected chain model with the \smc sampler and
show how we can perform the necessary steps using {\em Forward Filtering/Backward Sampling} (\ffbs)
\cite{carter1994gibbs, fruhwirth1994data}. For a more thorough description of \smc methods see \eg
\cite{DoucetJ:2011,doucet2001sequential}.

\subsection{Estimating the partition function using fully adapted \smc}
We propose to use a {\em fully adapted} \smc algorithm \cite{PittS:1999}.
That the sampler is \emph{fully adapted} means that the proposal distributions
for the resampling and propagation steps are optimally chosen with respect to minimizing the
variance of the importance weights, \ie the importance weights for a fully adapted sampler
are all equal. Using the optimal proposal distributions---which can significantly reduce the variance
of estimators derived from the sampler---is not tractable in general. However, as we shall see below,
this is in fact possible for the square lattice \pgm described above.

For the undirected chain model (see Figure~\ref{fig:hmmfig_b}), we let $\bar{\gamma}_k(\x_{1:k})$ be the \pmf induced by the sub-graph corresponding
to the first $k$ variables. 
Specifically, $\bar{\gamma}_k(\x_{1:k}) = \frac{\gamma_k(\x_{1:k})}{Z_k}$, where the unnormalized distributions $\gamma_k(\x_{1:k})$ are given by
\begin{subequations}
  \begin{align}
    \gamma_1 (\xk[1]) &= \bphi(\xk[1]),\\
    \gammak (\Xk) &= \gamma_{k-1}(\Xk[1:k-1]) \bphi(\xk) \bpsi(\xk,\xk[k-1]),    
  \end{align}
\end{subequations}
with $\bphi (\cdot), \bpsi (\cdot)$ as defined in~\eqref{eq:psiDef} and $Z_k$ being the normalizing constant for $\gamma_k(\x_{1:k})$.
We take the sequence of distributions $\bar\gamma_k(\x_{1:k})$ for $k = \range{1}{M}$ as the target distributions for the \smc sampler.
Note that $\bar{\gamma}_k(\x_{1:k})$ for $k < M$ is \emph{not}, in general, a marginal distribution under $p(X)$.
This is, however, not an issue since by construction $\bar\gamma_M(\x_{1:M}) = p(X)$ (where $\x_{1:M}$ identifies to $X$),
\ie at iteration $k=M$ we still recover the correct target distribution.

At iteration $k$, the \smc
sampler approximates $\bar{\gamma}_k(\x_{1:k})$ by a collection of particles
$\{\Xk^i\}_{i=1}^N$, where $\Xk = \{\x_1,\ldots, \xk \}$ is the set of all variables in column $1$
through $k$ of the \pgm. These samples define an empirical point-mass approximation of
the target distribution,
\begin{align*}
\widehat\gamma_k^N(\Xk) \eqdef \frac{1}{N}\sum_{i=1}^N  \delta(\Xk-\Xk^i),
\end{align*}
where $\delta(x)$ is the Kronecker delta. 
The standard \smc algorithm produces a collection of weighted particles. However, as mentioned above, in the fully adapted setting we use a specific choice of proposal distribution and resampling probabilities, resulting in
equally weighted particles \cite{PittS:1999}.

% Init
Consider first the initialization at iteration $k=1$. The auxiliary probability distribution $\bar\gamma_1(\x_1)$ corresponds to the \pgm induced by the first column of the original square lattice model. That is, the graphical model for $\bar\gamma_1(\x_1)$ is a chain (the first column of Figure~\ref{fig:hmmfig_a}). Consequently, we can sample from this distribution exactly, as well as compute the normalizing constant $Z_1$, using \ffbs. The details are given in the subsequent section. Simulating $\Np$ times from  $\bar\gamma(\x_1)$
results in an \emph{equally weighted} sample $\{\x_1^i\}_{i=1}^\Np$ approximating this distribution.

% Propagation
We proceed inductively and assume that we have at hand a sample $\{\Xk[1:k-1]^i\}_{i=1}^N$,
approximating $\bar{\gamma}_{k-1}(\Xk[1:k-1])$.  This sample is propagated forward by simulating,
conditionally independently given the particle generation up to iteration $k-1$, as follows: We
decide which particle among $\{\x_{1:k-1}^{j}\}_{j=1}^\Np$ that should be used to generate a new particle $\x_{1:k}^{i}$
(for each $i \in \set{1}{N}$) by drawing an {\em ancestor index} $a_k^i$ with probability
\begin{align}
  \label{eq:smc:resampling}
  \Prb(a_k^i = j) = \frac{ \nu_{k-1}^j }{ \sum_{l} \nu_{k-1}^l }, \qquad j \in \set{1}{N},
\end{align}
where $\nu_{k-1}^i$ are resampling weights. The variable $a_k^i$ is the index of the particle at iteration $k-1$ that will be
used to construct $\Xk^i$. 
Generating the ancestor indices corresponds to a selection---or
resampling---process that will put emphasis on the most likely particles.
This is a crucial step of the \smc sampler.
For the fully adapted sampler, the resampling weights $\nu_{k-1}^i = \nu_{k-1}(\Xk[k-1]^{i})$
are chosen in order to adapt the resampling to the \emph{consecutive target distribution} $\bar{\gamma}_k$ \cite{PittS:1999}.
Intuitively, a particle $\x_{1:k-1}^i$ that is probable under the marginal distribution $\sum_{\x_k}\bar\gamma_k(\x_{1:k})$
will be assigned a large weight. Specifically, in the fully adapted algorithm we pick the resampling weights according to
\begin{align}
\label{eq:smc:adjustmult}
\nu_{k-1}(\Xk[k-1]) = \sum_{\xk} \frac{\gammak (\Xk)}{\gamma_{k-1}(\Xk[1:k-1])}
= \sum_{\xk} \bphi(\xk) \bpsi(\xk,\xk[k-1]).
\end{align}

Given the ancestors, we simulate $\xk^i$
from the optimal proposal distribution: $\x_k^i \sim q(\cdot \mid \x_{k-1}^{a_k^i})$ for $i=\range{1}{\Np}$, where
\begin{align}
  \label{eq:smc:propagation}
  q(\x_k \mid \x_{k-1}) = \frac{\bphi(\xk)\bpsi(\xk,\xk[k-1])}{\sum_{\xk'} \bphi(\xk') \bpsi(\xk',\xk[k-1])}.
\end{align}
Again, simulating from this distribution, as well as computing the resampling weights \eqref{eq:smc:adjustmult},
can be done exactly by running \ffbs on the $k$th column of the model. Finally, we augment the particles as,
$\Xk^i \eqdef (\Xk[1:k-1]^{a_k^i},  \xk^i)$. As pointed out above, with the choices \eqref{eq:smc:adjustmult} and \eqref{eq:smc:propagation}
we obtain a collection of equally weighted particles $\{ \x_{1:k}^i \}_{i=1}^\Np$, approximating $\bar\gamma_k(\x_{1:k})$.

At iteration $k=M$, the \smc sampler provides a Monte Carlo approximation of the joint \pmf $p(X) = \bar\gamma_M(\x_{1:k})$.
While this can be of interest on its own, we are primarily interested in the normalizing constant $Z$ (\ie the partition function). However,
it turns out that the \smc algorithm in fact provides an estimator of $Z_k$ as a byproduct, given by
\begin{equation}
  \label{eq:smc:Zhat}
  \widehat Z_k^N \eqdef Z_1 \prod_{\ell = 1}^{k-1} \left( \frac{1}{N} \sum_{i=1}^N \nu_\ell^i \right).
\end{equation}
It may not be obvious to see why \eqref{eq:smc:Zhat} is a natural estimator of the normalizing constant $Z_k$.
However, it has been shown that this \smc-based estimator is unbiased for any $N \geq 1$ and $k=1,\ldots,K$.
This result is due to \cite[Proposition~7.4.1]{DelMoral:2004}.
Specifically, for our $2$-D constrained channel example, it follows that at the last iteration $k=M$ we have an unbiased estimator of the partition function 
\begin{align}
  \label{eq:smc:unbiased}
  \EE[\widehat Z_M^N] = Z.
\end{align}
Furthermore, under a weak integrability condition the estimator is asymptotically normal with a rate $\sqrt{\Np}$:
\begin{align}
  \label{eq:smc:clt}
  \sqrt{\Np}( \widehat Z_M^N - Z) \stackrel{d}{\rightarrow} \Normal(0, \sigma^2),
\end{align}
where an explicit expression for $\sigma^2$ is given in \cite[Proposition~9.4.1]{DelMoral:2004}.

\subsection{\smc samplers and Forward Filtering/Backward Sampling}
To implement the fully adapted \smc sampler
described above we are required to compute the sums involved in equations \eqref{eq:smc:adjustmult} and
\eqref{eq:smc:propagation}. By brute force calculation our method would be
computationally prohibitive as the complexity is exponential in the dimensionality $M$ of the chain. .
However, as we show below, it is possible to use \ffbs to efficiently carry out these summations.
This development is key to our proposed solution to the problem of estimating the partition
function, since the computational complexity of estimating the channel capacity is reduced from $\ordo (\Np M 2^M)$ (brute force) to
$\ordo (\Np M^2)$ (\ffbs).

Initially, at $k=1$, the graph describing the target distribution $\bar\gamma_1(\x_1)$
is trivially a chain which can be sampled from exactly by using \ffbs. Additionally, the normalizing constant
$Z_1$ can be computed in the forward pass of the \ffbs algorithm.
However, this is true for any consecutive iteration $k$ as well. Indeed, simulating $\x_k$ under $\bar\gamma_{k}$, conditionally on
$\Xk[1:k-1]$, again corresponds to doing inference on a chain. This means we can 
employ \ffbs to compute the resampling weights \eqref{eq:smc:adjustmult} (corresponding to a conditional normalizing constant computation)
and to simulate $\x_k$ from the optimal proposal \eqref{eq:smc:propagation}.

Let $k$ be a fixed iteration of the \smc algorithm. The forward filtering step of \ffbs is performed by sending messages
\begin{align}
\label{eq:forwardfiltering}
m^i_{j+1}(x_{j+1,k}) = \sum_{x_{j,k}} \psi (x_{j+1,k},x_{j,k}) \psi(x_{j,k}, x_{j,k-1}^i) m^i_j(x_{j,k}),
\end{align}
for $j=1,\ldots, M-2$, \ie from the top to the bottom of column~$k$. The resampling weights
are given as a byproduct from the message passing as
\begin{align}
\label{eq:ffbs:adjustmult}
\nu_{k-1}(\Xk[k-1]^i) &= \sum_{\xk} \bphi(\xk) \bpsi(\xk,\xk[k-1]^i) \nonumber \\
&= \sum_{x_{M,k}} \psi(x_{M,k}, x_{M,k-1}^i) m^i(x_{M-1,k}).
\end{align}
After sampling the ancestor indices $a_k^i$ as in \eqref{eq:smc:resampling}, we perform backward sampling
to sample the full column of variables $\xk$, one at a time in reverse order $j = M,\ldots,1$,
\begin{align}
\label{eq:backwardsampling}
x_{j,k}^i \sim \frac{\psi(x_{j,k},x_{j+1,k}^i) \psi(x_{j,k}, x_{j,k-1}^{a_k^i}) m_j^{a_k^i}(x_{j,k})}{\sum_{x_{j,k}'} \psi(x_{j,k}',x_{j+1,k}^i) \psi(x_{j,k}', x_{j,k-1}^{a_k^i}) m_j^{a_k^i}(x_{j,k}')},
\end{align}
with straightforward modifications for $j = 1$ and $M$. 
This results in a draw $\xk^i = \prange{x_{1,k}^i}{x_{M,k}^i}$ from the optimal proposal $q(\cdot \mid \x_{k-1}^{a_k^i})$ defined in \eqref{eq:smc:propagation}.
A summary of the resulting solution is provided in Algorithm~\ref{alg:smcffbs}.

\begin{algorithm}[htb]
   \caption{Channel capacity estimation}
   \label{alg:smcffbs}
\begin{algorithmic}
\STATE {\em Perform each step for $i = 1,\ldots,N$, except setting $\widehat Z_k^N$.}
\STATE Sample $\Xk[1]^i$ using \ffbs \eqref{eq:forwardfiltering}, \eqref{eq:backwardsampling}.
\STATE Set $\widehat Z_1^N = Z_1$.
\FOR{$k=2$ {\bfseries to} $M$}
\STATE Calculate $\nu_{k-1}(\x_{k-1}^i)$ using forward filtering \eqref{eq:forwardfiltering}-\eqref{eq:ffbs:adjustmult}.
\STATE Sample $a_k^{i}$ according to~\eqref{eq:smc:resampling}.
\STATE Sample $\xk^i$ using backward sampling \eqref{eq:backwardsampling}.
\STATE Set $\Xk^i = (\Xk[1:k-1]^{a_k^i}, \xk^i)$.
\STATE Set $\widehat Z_k^N = \widehat Z_{k-1}^N \left(\frac{1}{N} \sum_{i=1}^N \nu_{k-1}(\x_{k-1}^i)\right)$
\ENDFOR
\end{algorithmic}
\end{algorithm}

% ---------------------
% ONLY IN ARXIV VERSION
% ---------------------
\subsection{Practical implementation details}
For numerical stability it is important to use a few tricks in implementing
Algorithm~\ref{alg:smcffbs}. First, the size of the messages~\eqref{eq:forwardfiltering} grows quickly
with the chain dimension $M$ and the risk of overflow is big for realistic graph sizes. This can be resolved by instead working with the normalized messages
$\mu$, 
\begin{align}
  \label{eq:normalizedmessage}
  \mu_{j+1}^i(x_{j+1,k}) = \frac{1}{c_{j+1}^i} \sum_{x_{j,k}} \psi (x_{j+1,k},x_{j,k}) \psi(x_{j,k}, x_{j,k-1}^i) \mu_{j}^i(x_{j,k}),
\end{align}
where $c_{j+1}^i = \displaystyle \sum_{x_{j:j+1,k}} \psi (x_{j+1,k},x_{j,k}) \psi(x_{j,k},
x_{j,k-1}^i) \mu_j^i(x_{j,k})$ is just the normalization constant of the message. We can see that
using the normalized message $\mu_j^{i}$ instead of $m_j^i$ in \eqref{eq:backwardsampling} does not change the
distribution that we are sampling from. Furthermore, it is easy to verify that the resampling weights
are given by
\begin{align}
  \label{eq:efficient-implementation-weights}
  \nu_{k-1}(\Xk[k-1]^i) = \left(\prod_{j=1}^{M-2}c_{j+1}^i \right) \sum_{x_{M,k}} \psi(x_{M,k}, x_{M,k-1}^i) \mu^i(x_{M-1,k}).
\end{align}
Secondly, since we are actually interested in calculating the capacity, which is proportional to
$\log_2 Z$, we estimate the log-partition function as follows
\begin{align}
\log_2 \widehat Z_k^N = \log_2 \widehat Z_{k-1}^N + \log_2 \left( \sum_{i=1}^N \nu_{k-1}(\x_{k-1}^i)\right) - \log_2 \Np.
\end{align}
Note that in taking the logarithm of $\widehat Z_k^N$ we introduce a negative bias (\cf \eqref{eq:smc:unbiased}). However, the
estimator of the log-partition function (and thus also the capacity~\eqref{eq:bkg:capacity}) is nevertheless consistent and the bias decreases at a rate $\ordo(1/N)$.
Indeed, as we will see, in practice the bias is negligible and the error is dominated by the variance. 

Thirdly, in \smc implementations it is advisable to work with the logarithms of the
resampling weights. This will usually lead to increased numerical stability and help to
combat underflow/overflow issues.
With $\log_2 \nu_{k-1}(\Xk[k-1]^i)$ being the logarithm of \eqref{eq:efficient-implementation-weights}, we update the weights as:
\begin{subequations}
  \label{eq:num:adjustmult}
  \begin{align}
    %\nonumber
    % \log_2 \nu_{k-1}(\Xk[k-1]^i) &= \log_2 \sum_{x_{M,k}} \psi(x_{M,k}, x_{M,k-1}^i) \mu^i(x_{M-1,k})\\
    % &+ \sum_{j=1}^{M-2} \log_2 c_{j+1}^i, \\ 
    c &\gets \underset{i}{\text{max}} \left\{ \log_2 \nu_{k-1}(\Xk[k-1]^i) \right\},\\
    \label{eq:newadjustmult}
    \nu_{k-1}(\Xk[k-1]^i) &\gets 2^{\log_2 \nu_{k-1}(\Xk[k-1]^i) - c}.
  \end{align}
\end{subequations}
where $c$ is the maximimum of the log of the adjustment multipliers. 
Subtracting the maximum value $c$ from all the log-weights improves numerical stability and it does not change the
resampling probabilities \eqref{eq:smc:resampling} due to the normalization. However, we must add the constant $c$ to the
sequential estimate of the log-partition function, \ie
\begin{multline}
\label{eq:num:partition}
\log_2 \widehat Z_k^N = \log_2 \widehat Z_{k-1}^N + \log_2 \left( \sum_{i=1}^N \nu_{k-1}(\x_{k-1}^i)\right)
 \\- \log_2 \Np + c,
\end{multline}
where $\nu_{k-1}(\x_{k-1}^i)$ are the modified weights given by \eqref{eq:newadjustmult}.

%%% Local Variables:
%%% TeX-master: "smcIT.tex"
%%% End:

\section{Experiments}\label{sec:experiments}
We compare our algorithm to the state-of-the-art Monte Carlo approximation algorithm proposed in \cite{molkaraie2013information} on the same example that they consider as explained in Section \ref{sec:background}. Since the key enabler to the algorithm proposed in \cite{molkaraie2013information} is tree sampling according to \cite{HamzeF:2004}---a specific type of blocked Gibbs sampling---we will in the sequel refer to this algorithm as the {\em tree sampler}. All results are compared versus average wall-clock execution time. We run each algorithm $10$ times independently to estimate error bars as well as mean-squared-errors (MSE) compared to the true value (computed using a long run of the tree sampler).
For the \mcmc-based tree sampler, we use a burn-in of $10 \%$ of the generated samples when estimating the capacity.
The tree sampler actually gives two estimates of the capacity at each iteration; we use the average of these two when comparing to the \smc algorithm.

\begin{figure}[h!]
  \centering
    \includegraphics[width=0.45\textwidth]{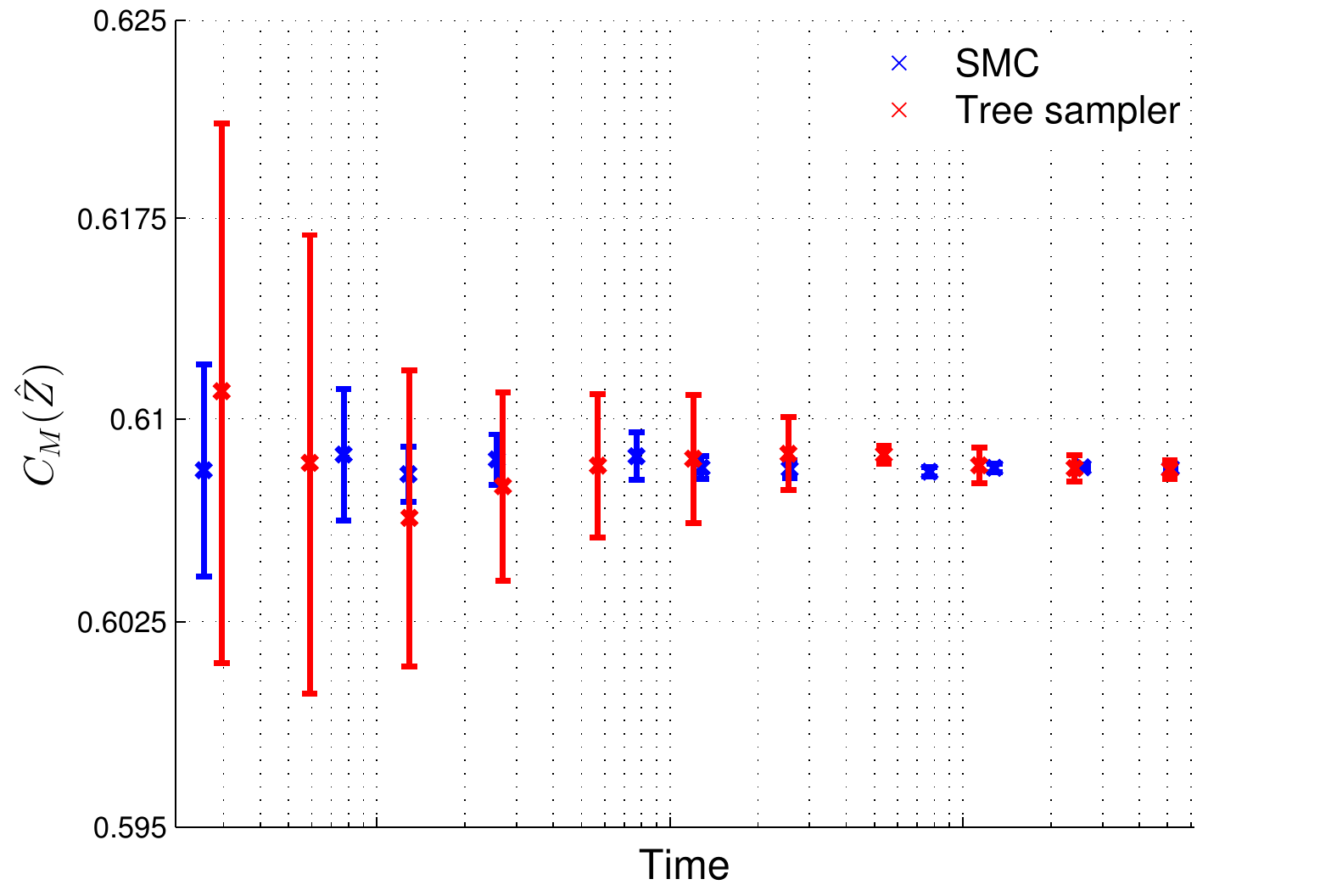}
    \caption{Estimates of the capacity $C_{10}$, with error bars, based on $10$ independet runs of our proposed \smc-based method and the tree sampler \cite{molkaraie2013information}. Plotted versus wall-clock time in log-scale. Note that this is also an upper bound on the infinite-size capacity, \ie $C_M \geq C_{\infty} \approx 0.5879$.}\label{fig:10x10est}
 \end{figure}

Consider first a channel with dimension $M=10$. We can see the results with error bars from $10$ independent runs in Figure~\ref{fig:10x10est} of both algorithms. The rightmost data point corresponds to approximately $20$k iterations/particles. Both algorithms converge to the value $C_{10} \approx 0.6082$. However, the \smc algorithm is clearly more efficient and with less error per fix computation time. We estimated the true value by running $10$ independent tree samplers for $100$k iterations, removed burn-in and taking the mean as our estimate. 

\begin{figure}[h!]
  \centering
    \includegraphics[width=0.45\textwidth]{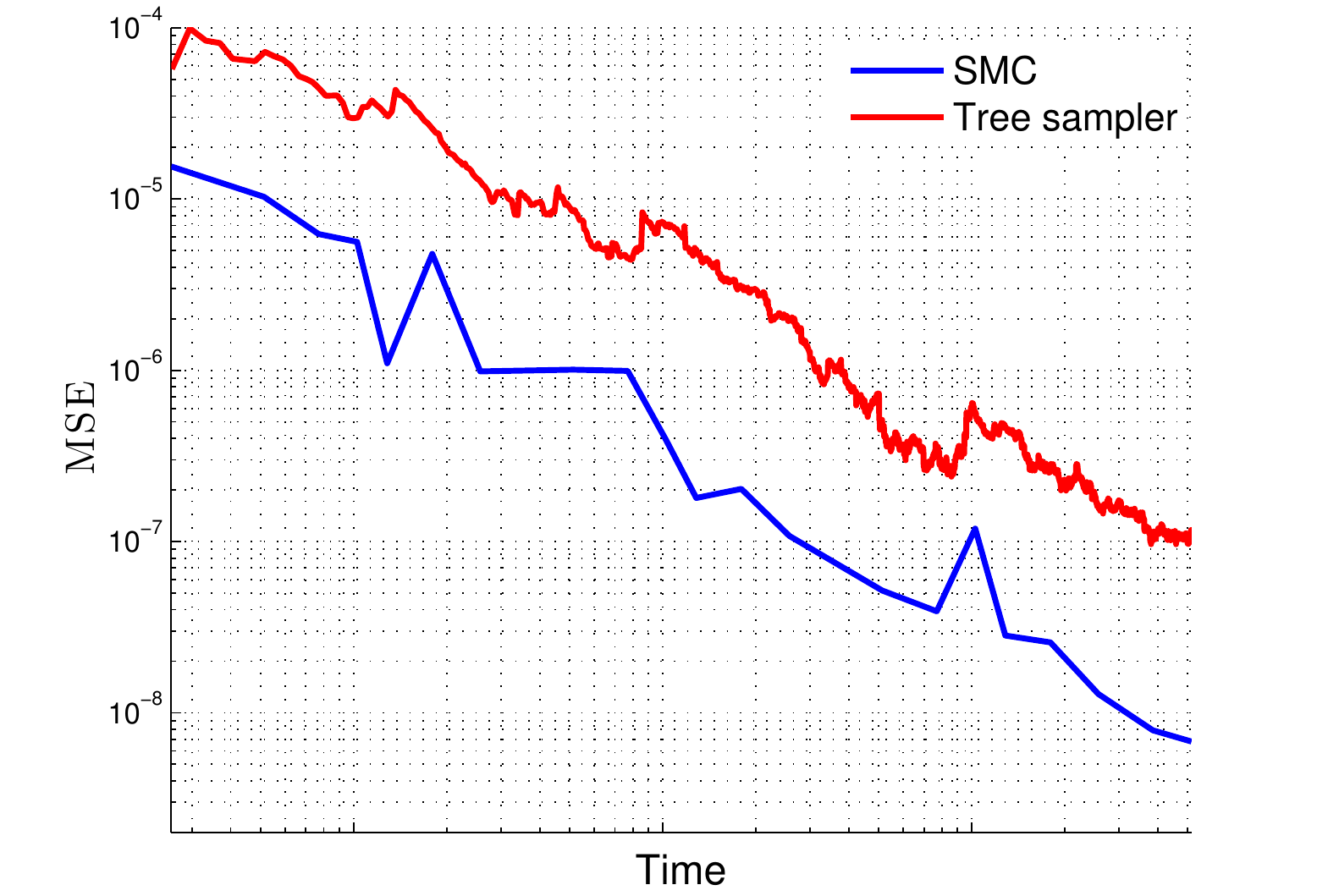}
    \caption{Mean-squared-error of the capacity $C_{10}$ estimates based on $10$ independet runs of our proposed \smc-based method and the tree sampler \cite{molkaraie2013information}. Plotted versus wall-clock time in log-log-scale.}\label{fig:10x10err}
\end{figure}

\begin{figure}[h!]
  \centering
    \includegraphics[width=0.45\textwidth]{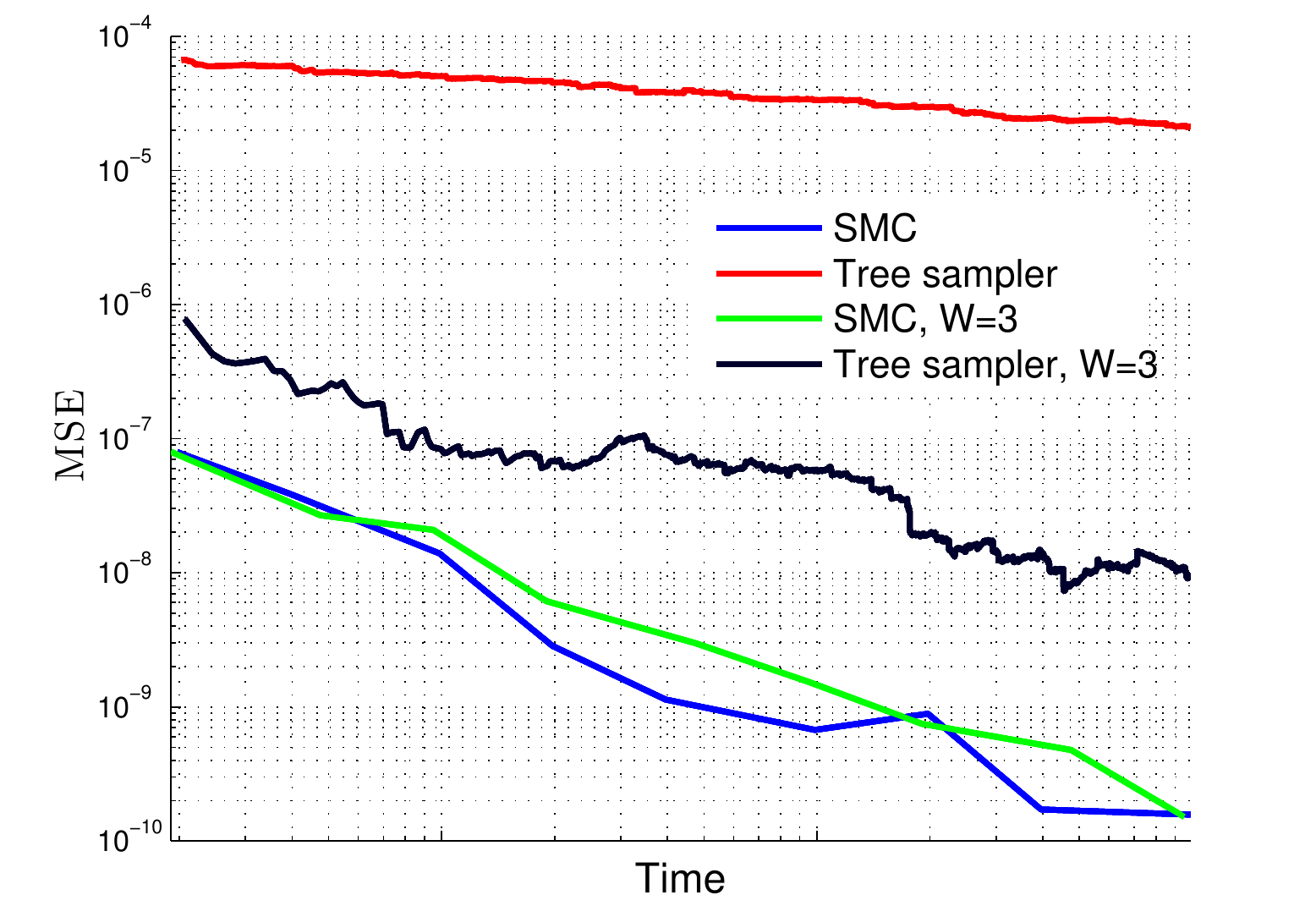}
    \caption{Mean-squared-error of the capacity $C_{60}$ estimates based on $10$ independet runs of our proposed \smc-based method and the tree sampler \cite{molkaraie2013information} for strip widths $1$ (standard) and $3$ respectively. Plotted versus wall-clock time in log-log-scale.}\label{fig:60x60err}
\end{figure}

The estimated true value was subsequently used to calculate the MSE as displayed in Figure~\ref{fig:10x10err}. The central limit theorem for the \smc sampler (see \eqref{eq:smc:clt}) tells us that the error should decrease at a rate of $1/N$ which is supported by this experiment. Furthermore, we can see that the \smc algorithm on average gives an order of magnitude more accurate estimate than the tree sampler per fix computation time.
In our second example we scale up the model to $M=60$, \ie a total of $3600$ nodes as opposed to $100$ in the previous example. The basic tree sampler performs poorly on this large model with very slow mixing and convergence. To remedy this problem \cite{molkaraie2013information} propose to aggregate every $W$ columns in the tree sampler and sample these exactly by simple enumeration, resulting in further blocking of the underlying Gibbs sampler. However, this results in an algorithm with a computational complexity exponential in $W$ \cite{molkaraie2013information}. The same strategy can be applied to our algorithm and we compare the tree sampler and \smc for widths $W=1$ and $3$.
There seems to be no gain in increasing the width higher than this for either method.
The resulting MSEs\footnote{For this model
the basic tree sampler converges too slowly and the tree sampler with $W=3$ was too computationally demanding to provide an accurate estimate
of the ``true'' value. For this reason, we estimate the true value by averaging $10$ independent runs of \smc with $N = 200$k.}
 based on $10$ independent runs of the tree sampler and the \smc algorithm are presented in Figure~\ref{fig:60x60err}.
As we can see the basic tree sampler converges very slowly, in line with results from \cite{molkaraie2013information}. On the other hand, our proposed \smc sampling method performs very well,
even with $W=1$, and on average it has more than an order-of-magnitude smaller error than the tree sampler with $W=3$. In comparing the two different \smc methods there seems to be no apparent gain in increasing the width of the strips added at each iteration in this case.

%%% Local Variables:
%%% TeX-master: "smcIT.tex"
%%% End:

\section{Conclusions}\label{sec:conclusions}
We have introduced an \smc method to compute the noiseless capacity of two-dimensional channel models. The proposed algorithm was shown to improve upon a
state-of-the-art Monte Carlo estimation method by more than an order-of-magnitude. Furthermore, while this improvement was obtained
using a sequential implementation, the \smc method is easily parallelizable over the particles
(which is not the case for the \mcmc-based tree sampler), offering further improvements by making use of modern computational architectures. This gain is of significant importance because the running time can be on the order of days for realistic scenarios.
Extensions to calculate the  information rate of noisy $2$-D source/channel
models by the method proposed in \cite{molkaraie2013information} are straightforward.

\section*{Acknowledgment}
% Supported by the project Probabilistic modeling of dynamical systems (Contract number:
% 621-2013-5524) funded by the Swedish Research Council, and
% the project Bayesian Tracking and Reasoning over Time (Reference: EP/K020153/1), funded by the EPSRC.
%The first author would also like to thank Lukas Bruderer for suggesting the application.
Supported by the projects \emph{Probabilistic modeling of dynamical systems} (Contract number:
621-2013-5524) and \emph{Learning of complex dynamical systems} (Contract number: 637-2014-466),
both funded by the Swedish Research Council.
%, and the project Bayesian Tracking and Reasoning over Time (Reference: EP/K020153/1), funded by the EPSRC.
We would also like to thank Lukas Bruderer for suggesting the application.

\bibliographystyle{IEEEtran}
\bibliography{IEEEabrv,smcITbib}

\end{document}